\documentstyle[11pt,aaspp4]{article}

\received{March 19 1998}
\accepted{May 4 1998}

\slugcomment{To be published in the Astrophysical Journal (Main Journal),
Vol. 505, October 1 1998 issue}

\lefthead{Wilson et al.}
\righthead{Obscuring `Torus' in NGC 2639}

\begin{document}

\title{The Ionization Fraction in the Obscuring `Torus' of an Active
Galactic Nucleus}

\author{A. S. Wilson\altaffilmark{1}}
\affil{Space Telescope Science Institute, 3700 San Martin Drive, Baltimore, MD
21218; awilson@stsci.edu}

\author{A. L. Roy, J. S. Ulvestad}
\affil{National Radio Astronomy Observatory, P. O. Box 0, Socorro,
NM 87801; aroy@aoc.nrao.edu, julvesta@aoc.nrao.edu}

\author{E. J. M. Colbert}
\affil{Mail Code 662, Laboratory for High Energy Astrophysics, NASA
Goddard Space Flight Center, Greenbelt,
MD 20771; colbert@lheamail.gsfc.nasa.gov}

\author{K. A. Weaver}
\affil{Johns Hopkins University, Department of Physics and Astronomy,
Baltimore, MD 21218; kim.weaver@pha.jhu.edu}

\author{J. A. Braatz}
\affil{National Radio Astronomy Observatory, P. O. Box 2, Green Bank,
WV 24944; jbraatz@nrao.edu}

\author{C. Henkel}
\affil{Max-Planck-Institut f{\"u}r Radioastronomie,
Auf dem H{\"u}gel 69, D-53121 Bonn, Germany; p220hen@mpifr-bonn.mpg.de}

\author{M. Matsuoka, S. Xue}
\affil{The Institute of Physical and Chemical Research (RIKEN), 
2-1, Hirosawa, Wako-shi, Saitama 351-01, Japan; matsuoka@postman.riken.go.jp,
xue@crab.riken.go.jp}

\author{N. Iyomoto}
\affil{Department of Physics, University of Tokyo,
3-1, Hongo 7-chome, Bunkyo-ku, Tokyo 113,
Japan; iyomoto@miranda.phys.s.u-tokyo.ac.jp}

\and

\author{K. Okada}
\affil{Institute of Space and Astronautical Science,
3-1-1, Yoshinodai, Sagamihara, Kanagawa 229, Japan; okada@astro.isas.ac.jp}


\altaffiltext{1}{Also Astronomy Department, University of Maryland,
College Park, MD 20742; wilson@astro.umd.edu}


\begin{abstract}

The LINER galaxy NGC 2639 contains a water vapor megamaser, suggesting the
presence of an edge-on nuclear
accretion disk or torus. This galaxy is thus
a good candidate for revealing absorption by the torus of any compact nuclear
continuum
emission. In this paper, we report
VLBA radio maps at three frequencies and an ASCA X-ray spectrum obtained to
search for free-free and photoelectric absorptions, respectively.
The radio observations reveal a compact
($<$ 0.2 pc) nuclear source with a spectrum that turns over sharply near
5 GHz. This turnover may reflect either synchrotron self-absorption or
free-free absorption. The galaxy is detected by ASCA with an observed
luminosity of $1.4 \times 10^{41}$ erg s$^{-1}$ in the 0.6 -- 10 keV band.
The X-ray spectrum shows emission in excess of a
power-law model at energies greater than 4 keV; we interpret this excess
as compact, nuclear, hard X-ray emission with the lower energies 
photoelectrically absorbed by an
equivalent hydrogen column of $\simeq$ 5 $\times$ 10$^{23}$ cm$^{-2}$.

If we assume that 
the turnover in the
radio spectrum is caused by free-free absorption
and that both the free-free and photoelectric absorptions are produced
by the same gaseous component, the ratio
$\int n_{e}^{2} dl/\int n_{H} dl$ may be determined. 
If the masing molecular gas is responsible for both
absorptions, the required ionization fraction is
$\gtrsim 1.3 \times 10^{-5}$, which is comparable to the theoretical
{\it upper} limit derived by Neufeld, Maloney \& Conger (1994) for X-ray
heated molecular gas. The two values may be reconciled if the molecular gas
is very dense -- $n_{H_{2}} \gtrsim 10^{9}$ cm$^{-3}$.
The measured ionization fraction is also consistent with the idea
that both absorptions occur in a hot ($\sim$ 6,000K),
weakly ionized (ionization fraction a few times 10$^{-2}$) atomic region that
may co-exist with the warm molecular gas. If this is the case, the absorbing
gas is $\sim$ 1 pc from the nucleus. We rule out the possibility that both
absorptions occur in a fully ionized gas near 10$^{4}$K. 
If our line of sight passes through more than one phase, the atomic gas probably
dominates the free-free absorption while the molecular gas may dominate
the photoelectric absorption.
\end{abstract}




%

\newpage
\section{Introduction}

Over the last decade, it has become clear that many, perhaps all, active
galactic nuclei, are surrounded by dusty accretion disks or tori on the pc
or sub-pc scale (e.g. Antonucci 1993). 
When viewed equatorially, these tori hide emission from 
the central regions and they are believed to be responsible for the
observational difference between broad-line objects (Seyfert 1s, broad-line
radio galaxies and quasars), in which the torus is viewed near pole-on,
and narrow-line objects (Seyfert 2s, narrow-line radio galaxies), in which the 
torus is viewed near edge-on. This model is supported by a wide range
of observational results on narrow-line objects -- polarized
broad lines (e.g. Tran 1995), reddened nuclei at optical wavelengths
(Mulchaey, Wilson \& Tsvetanov 1996), broad infrared 
recombination lines (e.g. Goodrich, Veilleux \& Hill 1994), large gas
columns to the nuclei as inferred from photoelectric absorption of soft X-rays
(e.g. Turner et al. 1997) and bi-cones of ionized gas aligned with the
radio ejecta (e.g. Wilson \& Tsvetanov 1994).

The obscuring material, which appears to take the form of
gas clouds in a geometrically thick torus
or a warped, thin disk, is illuminated by the central UV and X-ray continuum
source. Calculations of the physical and chemical properties of
the gas have been made
by Krolik \& Lepp
(1989) and Neufeld, Maloney \& Conger (1994, hereafter NMC).
The UV and soft X-rays are 
expected to be absorbed in thin layers at the surface of the clouds, but
hard X-rays may penetrate and heat the interiors. This X-ray heated gas may
possess a two phase structure, in which an atomic phase at $T \approx$
5,000 -- 8,000K coexists with a molecular phase at $T \approx$ 600 -- 2,500 K
(e.g. NMC). The ionization fraction is expected to
be $\lesssim$ 10$^{-5}$ in the molecular region
and a few times 10$^{-2}$ in the atomic region, but the exact
values are sensitive
to the X-ray flux and gas density and to details of the chemical and physical
processes in the gas. Clearly, an observational determination of the
ionization fraction would be of value and is the goal of the present paper.

The ionization fraction may be estimated, in principle, by comparing the
emission measure, $\int n_{e}^{2} dl$
(from measurements of free-free absorption of the nuclear radio emission)
with the
total equivalent
hydrogen column, $\int n_{H} dl$ (from measurements of
photoelectric absorption of nuclear X-rays),
both through the disk or torus.
To provide maximum path length for these absorptions, the disk needs to
be oriented close to edge-on. Studies of H$_{2}$O megamaser emission from the 
nucleus of NGC 4258 reveal that the maser emission arises in a thin Keplerian
accretion 
disk which is viewed very 
close to edge-on (Watson \& Wallin 1994; Miyoshi et al.
1995). More recent VLBI mapping of other H$_{2}$O megamasers shows that, in
almost all cases, the maser emission traces a line on the sky
with kinematics consistent with an edge-on, rotating disk 
(Greenhill \& Gwinn 1997; Greenhill, Moran \& Herrnstein 1997;
Trotter et al. 1998).
Thus galaxies with detected H$_{2}$O megamaser emission provide
the best opportunity for measuring the ionization fraction of the
circumnuclear disk or torus.

NGC 2639 has a LINER-type nucleus
(e.g. Ho, Filippenko \& Sargent 1993) and H$_{2}$O megamaser emission
(Braatz, Wilson \& Henkel 1994). Although
a VLBA map of the maser emission is not yet available, the systemic maser
emission has been found to drift redward at a similar rate to that
seen in NGC 4258 (Wilson, Braatz \& Henkel 1995). In NGC 4258, this redward
drift is known to result from the centripetal acceleration of clumps of
masing gas on the
near side of the edge-on accretion 
disk, as the gas passes in front of the nuclear
radio source (Herrnstein et al. 1997). By analogy with NGC 4258, it may
be argued that the redward drift in NGC 2639 arises in the same way and that
it too contains an accretion disk viewed very close to edge-on. We
therefore chose NGC 2639 for a search for absorption by the putative disk.

Throughout this paper, we adopt a velocity of NGC 2639 with respect to the 
microwave background radiation of V$_{3K}$ = 3,434 km s$^{-1}$ (de
Vaucouleurs et al. 1991) and
a Hubble constant of 75 km s$^{-1}$ Mpc$^{-1}$, giving a distance
of 45.8 Mpc and a scale of 222 pc per arc second.

\section{Observations}
\subsection{VLBA Radio Observations}

NGC 2639 was observed as part of our survey of Seyfert galaxies on
May 31, 1996, using the ten antennas of the
VLBA\footnote{The VLBA is part of the National Radio Astronomy
Observatory, a facility of the National Science Foundation,
operated under cooperative agreement by Associated
Universities, Inc} at 1.7, 5.0, and 15
GHz.  Baseline lengths were from 130 to 5000 km, and NGC 2639 was
observed for 57 min at 1.7 GHz, 48 min at 5.0 GHz, and 46 min at 15
GHz.  We recorded left circular polarization with 32-MHz bandwidth and
2-bit sampling.  Calibration and imaging were done with the AIPS
software, using standard methods.  The flux density scale was
calibrated using standard VLBA antenna gains and measurements of
$T_{\mathrm{sys}}$ made every 1 -- 2 minutes.

We phase referenced to the nearby source J0832$+$4913 ($2.1^\circ$
away, adopted J2000 position 
RA $=08^{h}$ $32^{m}$ $23^{s}.21671$,
Dec = $49^{\circ}$ $13'$ $21''.0388$ $\pm 0.5$
mas [Eubanks 1995, private communication]),
with cycle times of 9 min at 1.7 GHz and 5.0 GHz, and 2 min at
15 GHz.  
After phase referencing, the data were imaged, deconvolved, and self
calibrated, with convergence achieved after two iterations of
phase-only self calibration. This yielded high-resolution images (from
untapered
u-v data)
with FWHM beam sizes of
6.4 mas $\times$ 4.6 mas in p.a. 159$^{\circ}$ at 1.7 GHz,
2.0 mas $\times$ 1.7 mas in p.a. 169$^{\circ}$ at 5 GHz,
and 1.2 mas $\times$ 0.75 mas in p.a. 105$^{\circ}$ at 15 GHz.
The r.m.s. noise on these images is $n$ = 0.14, 0.35 and 0.26 mJy
(beam area)$^{-1}$ at 1.4, 5.0 and 15 GHz, respectively, which is similar to
the expected thermal noise at 1.4 and 15 GHz, but somewhat higher than
the expected value (0.15 mJy (beam area)$^{-1}$) at 5.0 GHz.
For
measuring accurate spectral indices, we tapered the array at 5
and 15 GHz to produce approximately
the same beamwidth as for the untapered 1.7-GHz
array, using spacings from 0.5, 1.2 and 4.2 M$\lambda$ to
40 M$\lambda$ at 1.7, 5, and 15 GHz, respectively.
A restoring
beam of 6.4~mas $\times$ 4.6 mas (FWHM) was then used at all three
frequencies. 
The rms noise levels in the
tapered images were 0.69 mJy (beam area)$^{-1}$ at 5 GHz and
0.77 mJy (beam area)$^{-1}$ at 15 GHz.

The flux-density measurements have uncertainties due to flux-scale
calibration, thermal noise, and deconvolution errors.
The flux scale has a nominal uncertainty of 5\% at each
frequency (uncertainties quoted throughout this paper are r.m.s.).
Deconvolution effects were estimated approximately by
experimenting with different depths of cleaning and sizes of the clean
boxes. The test was performed on NGC 1068, which was observed during the
same run and in the same way as NGC 2639.
A grid of nine different sets of reasonable parameters
produced a distribution of total flux densities that had a standard
deviation of 6.8\%, and we adopt this estimate as the error induced
by deconvolution.
Resolution effects should be minimal in the measurement of
spectral indices because we have matched the beamwidths at all
frequencies and because the source is only slightly resolved.
Combining all the above effects in quadrature, the uncertainty on the
flux-density measurements is
$\sqrt{(0.08 S)^{2} + n^{2}}$ mJy, where $S$ is the flux density and $n$ is the
noise given above for each frequency.

\subsection{ASCA X-ray Observations}

NGC 2639 was observed by $ASCA$ (Tanaka, Inoue \&
Holt 1994) on 16 April, 1997.  $ASCA$ carries two 
pairs of instruments known as the solid-state
imaging spectrometers (SIS; hereafter S0 and S1) and the 
gas-imaging spectrometers (GIS; hereafter G2 and G3).
The SIS data were obtained in 1-CCD mode, converted 
to BRIGHT mode, and hot and flickering pixels were removed. 
Good time intervals for the SIS and GIS were
determined following Weaver et al.\ (1994), with 
additional SIS screening to remove   
times $\le$60 s after South Atlantic Anomaly passages and transitions 
from satellite day to satellite night.

Data were extracted from within circular regions centered
on the source with radii $2.3'$ and $5.3'$ in the SIS
and GIS, respectively (larger source regions for the
SIS do not improve the signal).  The SIS background was 
obtained from an annular ring of width $\sim1.5'$ on the source
chip while the GIS background was obtained from circular 
regions in the source field chosen to avoid a nearby
serendipitous X-ray source.  NGC 2639 has a total $ASCA$ 
count rate of 0.01 to 0.02 counts s$^{-1}$ depending on 
the detector, 70\% to 81\% of which is made up of 
background counts.  Subtracting this
background yields maximum count rates of 0.01 and 0.006
counts s$^{-1}$ (keV)$^{-1}$ for S0 and G3 respectively.
We measure smaller count rates in S1 and G2, in which
NGC 2639 is positioned about $8'$  
from the optical axis.  The effects of vignetting  
cause the galaxy to be indistinguishable from the 
background in these detectors above $\sim3$ keV.
In S0 and G3, NGC 2639 is
$\sim5'$ from the optical axis and thus less 
affected by vignetting.  Therefore we consider only 
data from S0 and G3 in our spectral fits.

For a weak source like NGC 2639, incorrect background subtraction
can cause spurious
results. Therefore we have examined background
taken from blank-sky fields in addition to background
taken from near the source (described above).
The choice
of background makes little difference for S0, but has a
significant effect on the data from G3.  We therefore performed
two sets of spectral fits with the different
G3 backgrounds. The qualitative features of the spectrum do
not depend on which background is used and the errors quoted
(Section 3.2 and Table 2) include the uncertainty in the background.

\section{Results}
\subsection{Radio Structure and Spectrum}

Radio fluxes of the nuclear source and beam sizes are given in Table 1 while
the radio spectrum is shown in Fig. 1. Our 5 GHz flux of 47 $\pm$ 4 mJy is 
higher than
both a previous VLBI measurement of 27 $\pm$ 4 mJy (Hummel et al.
1982, made between June 1980 and April 1981) and
the peak brightness 
of 23.4 $\pm$ 0.9 mJy beam$^{-1}$ in a VLA map (Ulvestad \& Wilson
1989, made on Feb 3 -- 4 1985),
indicating the source is variable.
The source is unresolved at 1.7 and 
5.0 GHz,
but extended at 15 GHz. 
The peak flux density in our highest resolution 15 GHz map is 33.8 mJy
(beam area)$^{-1}$, significantly lower than the total
flux of 45 $\pm$ 4 mJy. The
FWHM deconvolved source
size is 0.70 $\times$ $<$ 0.15 mas
(0.16 $\times$ $<$ 0.03 pc) with major axis in p.a. 111$^{\circ}$.
For comparison, VLA observations reveal an incompletely resolved triple source
with overall length 1.4$^{\prime\prime}$ (310 pc) in p.a. 105$^{\circ}$
(Ulvestad \& Wilson 1989).
The agreement between the p.a.'s of the sub-pc
scale and the hundreds of pc scale radio emission strongly suggests that the
extension found with the VLBA observations represents the inner part of
the jet presumed to fuel the VLA-observed structure.
The spectrum of the nuclear source (Fig. 1) is flat between 5 and 15 GHz
($\alpha^{15}_{5} = +0.04 \pm 0.08$, $S \propto \nu^{-\alpha}$) and turns over
sharply below
5 GHz ($\alpha^{5}_{1.7} = -1.8 \pm 0.1$). We now discuss various
interpretations of
the emission mechanism of the nuclear radio source and the cause of the low
frequency turnover.

The brightness temperatures, $T_{b}$, of the nuclear source are
$>$ 1.5 $\times$ 10$^{8}$K, $>$ 9.8 $\times$ 10$^{8}$K
and $>$ 2.9 $\times$
10$^{9}$K at 1.7, 5.0 and 15 GHz, respectively. For comparison, the brightness
temperature of the source S1 in NGC 1068, which is believed to coincide with
the nucleus
and may be thermal in origin, is 
of order a few $\times 10^{5}$ K to a few $\times
10^{6}$ K at 8.4 GHz (Gallimore et al. 1997; Roy et al. 1998). The high
brightness temperature of the source in NGC 2639 makes a thermal origin
unlikely.

The three processes that could, in principle, be
responsible for the low frequency turnover
are synchrotron self-absorption, Razin-Tsytovich suppression
and free-free absorption, and we discuss each in turn. We can ascribe
an effective temperature $T_{e}$ to electrons of a given energy through the
equation $3 k T_{e} = \gamma_{e} m_{e} c^{2}$ or
$T_{e} = 2.0 \times 10^{9} \gamma_{e}$K, where $\gamma_{e}$ is the relativistic
gamma factor of the radiating electrons, $m_{e}$ is the electron mass, $c$
is the speed of light and $k$ is Boltzmann's constant (e.g. Longair 1981).
For a synchrotron
self-absorbed source, $T_{b} = T_{e}$. The brightness temperature measurements
are only lower limits and are obviously consistent with a
synchrotron self-absorption interpretation. The source is, however, resolved
in one dimension
at 15 GHz. If we adopt a 15 GHz size of 0.16 $\times$ 0.03 pc (where the
minor axis extent is taken to be equal to its upper limit) and assume
the size is the same at 5 GHz (where the spectrum begins to turn over), then
$T_{b}$(5 GHz) $\simeq 2.9 \times$ 10$^{10}$K, which allows the
possibility of
synchrotron self-absorption. The mean magnetic field
strength would be $\sim$ 0.5
gauss (using equation (1) of Kellermann \& Pauliny-Toth
1981), which is similar to the equipartition value of $\sim$ 0.3 gauss
(calculated assuming that the total energy in cosmic rays is twice that
in electrons, the radio source is optically thin at 15 GHz, the electron
energy spectrum 
extends from radiated frequencies 10 MHz to 100 GHz with index $\gamma$ = 2.5
($N(E) \propto E^{-\gamma}$), and
that the source is ellipsoidal in shape with axes
0.16 $\times$ 0.03 $\times$ 0.03 pc).
We should, however, bear in mind that the resolved flux at 15 GHz is only
$\simeq$ 25\% of the total flux, so most of the observed emission may come from
unresolved regions with higher brightness and weaker magnetic fields.

Razin-Tsytovich suppression occurs at frequencies below
$\simeq 20n_{e}/B_{\perp}$,
where $n_{e}$ is the thermal 
electron density (in cm$^{-3}$) within the synchrotron-emitting region
and $B_{\perp}$ is the 
perpendicular component of the magnetic
field (in gauss). Comparison with the equation for free-free absorption
(see below) shows that the Razin-Tsytovich cut-off frequency will lie 
above the free-free absorption frequency only if
$B_{\perp} \le 4 \times 10^{-5} L^{-1/2}$ gauss, where $L$ is the dimension of
the absorbing region in pc (e.g. Moffet 1975). We can take
the measured maximum extent of the source at 15 GHz -- 0.16 pc -- as an upper 
limit for $L$, so for Razin-Tsytovich suppression to occur at a higher frequency
than
free-free absorption, the magnetic field would need to be
$B_{\perp} \le 1 \times 10^{-4}$ gauss, which is
more than three orders of magnitude below
equipartition. 
We conclude that Razin-Tsytovich suppression is an implausible explanation for
the low-frequency turnover in NGC 2639.

To investigate the possibility that free-free absorption
by ionized gas along the line of sight to the radio source is responsible for 
the low frequency cut-off, we assume the observed spectrum may be described
by:
$$S(\nu) = a \nu^{-\alpha} e^{-\tau(\nu)} \eqno(1)$$
with
$$\tau(\nu) = 8.235 \times 10^{-2} T_{e}^{-1.35} \nu^{-2.1} \int n_{e}^{2} dl
\eqno(2) $$
where $a$ is a constant,
$\tau(\nu)$ is the optical depth to free-free absorption, $T_{e}$ (K)
and $n_{e}$ (cm$^{-3}$) are the
electron temperature and density of the absorbing gas, $\nu$ is the
frequency (GHz)
and $l$ is the path length in pc (e.g. Mezger \& Henderson 1967).
The three observed flux densities may be used to solve for the three
unknowns ($a$, $\alpha$ and $\tau$) in equation (1),
given the known dependence of $\tau$ on $\nu$ in
equation (2). The result for the spectral index and optical depth is 
$\alpha$ = 0.25 and
$\tau$(5 GHz) = 0.25. The value of $\alpha$ should not be taken too seriously
(a flux measurement at a higher frequency is needed to tie down the
optically thin part of the spectrum),
but $\tau$ is well defined by the observed spectrum.
The spectrum from this model is shown by the line in Fig. 1.
The derived value of $\tau(\nu)$ implies
$$\int n_{e}^{2} dl = 2.3 \times 10^{7} T_{e4}^{1.35} \mathrm{cm}^{-6} pc
\eqno(3)$$
where $T_{e4} = T_{e}/10^{4}$ K.
We discuss the nature of the absorbing gas in Section 4. 

\subsection{X-ray Spectrum}

We first modelled the X-ray spectrum as a single, uniformly-absorbed power-law,
but found the fit to be poor (model A, Table 2), with the data
showing an excess over the model above 4 keV. This excess emission is 
independent of the choice of G3 background and probably
represents a heavily absorbed, hard component. A model (model B)
comprising a single
power-law spectrum absorbed by an intrinsic medium covering 93\% of the source
(i.e. ``partial covering'') and the
Galactic column density (N$_{H}$ = $3\times10^{20}$ cm$^{-2}$)
covering the whole source, provides
an excellent description of the data (again, independent of the 
choice of G3 background). The part of the X-ray source covered by only
the Galactic column could be scattered or intrinsically extended X-ray
emission. The best estimate of the photon index is $\Gamma$ = 2.4 $\pm$ 0.4.
Constraining the power-law index of the photon spectrum
to $\Gamma$ = 1.9 (close to the `canonical'
value for Seyfert galaxies) does not significantly worsen the fit
(model C).  We also tried a model that consists of a power
law absorbed by the Galactic column plus a Gaussian function
to represent the Fe K$\alpha$ 
emission often seen in active galaxies (model D).
Such a model 
describes the high-energy 
excess as well as models B and C; however, the line energy   
is much too low to be Fe K$\alpha$ emission at the redshift 
of NGC 2639 (Table 2). 

Both models B and C require an intrinsic column density
$\int n_{H} dl = (4 - 5) \times 10^{23}$ cm$^{-2}$,
although columns half to almost three times 
as large are acceptable at the 90\% confidence level (Table 2).
The observed S0 and G3 spectra 
and model B folded through the instrumental response are shown
in Figure 2.

\section{The Ionization Fraction}

A number of assumptions must be made in calculating the ionization 
fraction. First, we assume that the low frequency turnover in the radio
spectrum results from free-free absorption.
There is now strong evidence that the VLBI-scale radio sources in several
Seyfert galaxies suffer free-free absorption at GHz frequencies
(Halkides et al. 1998; Roy et al. 1998; Ulvestad et al. 1998), and our spectrum
of NGC 2639 is consistent with this process (synchrotron
self-absorption is, however, a viable alternative - see Section 3.1).
Second, the lines of sight to the hard X-ray source and the compact radio 
source must be the same. While the resolution of the VLBA is high and the
radio source is compact ($<$ 0.2 pc), coincidence of the two sight lines is
not guaranteed. For example, if some of the radio emission originates from
a compact jet, it may be spatially separated from the presumed nuclear
hard X-ray
source. Third, we assume that the same gas is responsible for both
free-free and photoelectric absorptions. This is not necessarily so,
even if the two lines of sight coincide: it is possible
that the photoelectric absorption is dominated by weakly ionized atomic or
molecular gas which is dense or
far from the central ionizing source, while the
free-free absorption may
result from highly ionized gas of low density or 
close to this source. The apparent ionization fraction would then 
represent a suitably weighted average along the
line of sight and is not necessarily representative of 
this parameter
in a particular gaseous component. Later in this Section (Section 4.4),
we explore the
consequences of abandoning this third assumption.

Our measurements imply
$\int n_{e}^{2} dl/\int n_{H} dl$ = 140 $T_{e4}^{1.35}$ cm$^{-3}$,
with a formal uncertainty of a factor of $\sim$ 2 (dominated by
the uncertainty in $\int n_{H} dl$). For a
uniform absorbing medium, the ionization fraction is then:
$$x_{e} = n_{e}/n_{H} = 1.2 \times 10^{-4} T_{e4}^{0.675} n_{H10}^{-0.5}
\eqno(4) $$
where $n_{H10} = n_{H}/10^{10}$ cm$^{-3}$. The uncertainty in
$\int n_{H} dl$ translates into an uncertainty of $\sim$ 1.4 in $x_{e}$.
Within the context of a model in which dense gas is heated and photoionized by a
nuclear UV and X-ray source, the absorption could occur in a warm, weakly
ionized molecular phase, a hot, weakly ionized atomic phase or a hot, 
highly ionized atomic phase (listed in order of decreasing distance from the
ionizing source). We consider these three possibilities in turn.

\subsection{Warm Molecular Phase}

For water vapor maser emission,
we require (e.g. Elitzur, Hollenbach \& McKee 1989)
$n_{H_{2}} \lesssim 10^{10}$ cm$^{-3}$ (to avoid collisional
quenching)
and gas temperatures $T_{g} \sim 600$K (for collisional pumping). Thus, if
the observed column of absorbing gas comprises the masing region,
equation (4) implies
$x_{e} \gtrsim 1.3 \times 10^{-5}$ (with the equality for 
$n_{H_{2}} = 10^{10}$ cm$^{-3}$), assuming $T_{e} = T_{g}$. This {\it lower}
limit
is comparable with the {\it upper} limit of $x_{e} \lesssim 10^{-5}$
derived theoretically for X-ray heated molecular gas by NMC. 
The theoretical,
column-averaged ionization fraction may be as high as several $\times$
10$^{-5}$, depending on the X-ray flux, gas column and pressure (P. R. Maloney,
private communication). Thus 
warm molecular gas 
could potentially provide both the 
free-free and photoelectric absorptions, but only if the density is
high ($n_{H_{2}} \gtrsim 10^{9}$ cm$^{-3}$).

\subsection{Hot, Weakly Ionized, Atomic Phase}

If, on the other hand, the
absorption occurs in the hot ($\sim$ 6,000K) atomic region in pressure
balance with the
molecular region (e.g. NMC), equation (4)
implies $x_{e} \gtrsim 2.7 \times 10^{-4}$,
which is consistent with the few times 10$^{-2}$ expected by NMC.
If our line of sight passes through this atomic region, its free-free
absorption should dominate the molecular phase (NMC). 
Demanding NMC's ionization fraction of $\sim 2 \times 10^{-2}$, our data imply
$n_{H} \sim 2 \times 10^{5}$ cm$^{-3}$ in the atomic region and a path length
of $N_{H}/n_{H} \sim 1$ pc for the absorbing gas.

The intrinsic 1 -- 100 keV X-ray luminosity of NGC 2639 is
$L_{X} = 1.5 \times 10^{42}$ erg s$^{-1}$ (for $\Gamma = 1.9$)
or $1.1 \times 10^{42}$ erg s$^{-1}$ (for $\Gamma = 2.4$). The X-ray
flux incident on the gas is $F_{X} = 10^{5}F_{5}$ erg cm$^{-2}$ s$^{-1}$,
where $F_{5} = 0.084 (L_{X}/10^{42}$ erg s$^{-1}$) $(R/$pc)$^{-2}$ and
$R$ is the distance from the X-ray source to the illuminated surface of the
gas. For the masing molecular gas and the absorbing atomic phase
to co-exist, NMC's Fig. 1 implies
$F_{5}/(N_{24}^{0.9} p_{11}) \simeq 10$,
where $N_{H} = 10^{24} N_{24}$ cm$^{-2}$ and
$p/k = 10^{11}p_{11}$ K cm$^{-3}$ ($p$ is the pressure). 
Taking $N_{24} = 0.5$ (from the X-ray absorption),
$n_{H} = 2 \times 10^{5}$ cm$^{-3}$ (see above) and
T = 6,000K for the atomic region, we find
$F_{5} \simeq 0.06$ and $R \simeq 1.2-1.4$ pc. This size scale corresponds to
$\simeq 5-6$ mas, 
which will be easily resolvable in planned VLBA observations of
the H$_{2}$O megamaser.

\subsection{Hot, Fully Ionized, Atomic Phase}

Next we investigate the hypothesis that a fully ionized ($x_{e} \simeq 1$)
gas at 10$^{4}$K is responsible for both absorptions. In this case,
equation (4) gives $n_{e} \sim n_{H} \sim 140$ cm$^{-3}$ and the absorber
path length $l$ = $N_{H}/n_{H} \sim 1$ kpc. For a Str\"omgren sphere
geometry, the required number of ionizing photons is
$$N_{*} = \alpha_{B} n_{e}^{2} 4 \pi l^{3}/3 \eqno(5) $$ 
($\alpha_{B}$ is the total hydrogen recombination coefficient in case B),
giving $N_{*} = 9.6 \times 10^{56}$ photons s$^{-1}$. This rate of ionizing
photons may be compared with 
$N_{*} = 3.3 \times 10^{53}$ photons s$^{-1}$ obtained by extrapolating the
measured hard X-ray flux down to
the Lyman limit with $\Gamma = 2.4$, the best estimate of the photon index
of the hard X-ray source. This 
value of $\Gamma$ is also typical of radio-quiet quasars between the 
Lyman limit and 1 keV (e.g. Laor et al. 1997); however, the 
spectra of LINERs in the ionizing UV are
poorly known, so our estimate is very uncertain. Nevertheless, given that the
required value of $N_{*}$ is more than three
orders of magnitude larger than this estimated value,
we rule out the possibility that both
free-free and photoelectric absorptions occur in the fully ionized phase.

\subsection{Multiple Phases}

Lastly, we abandon the assumption that the same gas is responsible for both
types of absorption and consider the possibility that the free-free absorption 
occurs
in the fully ionized gas at 10$^{4}$K while the photoelectric absorption is
dominated by another, weakly ionized component. Taking
$N_{*} = 3.3 \times 10^{53}$ photons s$^{-1}$ from the extrapolation
of the X-ray spectrum, $T_{e4}$ = 1 and combining equations (3) and (5), we
have $n_{e} \sim 1,000$ cm$^{-3}$ and $l \sim 20$ pc, suggestive of a large,
dense cloud in the narrow line region. The implied column density 
through the fully
ionized layer is $N_{H} \simeq 7 \times 10^{22}$ cm$^{-2}$, a factor of 7
lower than inferred from the X-ray spectrum, so a weakly
ionized column of $\sim 4 \times$ 10$^{23}$ cm$^{-2}$ would also have to be
present. The fully ionized gas would produce an H$\beta$ flux of
1.2$\times$10$^{-12}$ erg cm$^{-2}$ s$^{-1}$, which is factor of 8 larger than
the observed, reddening-corrected H$\beta$ flux from the nucleus of NGC 2639
(Ho, Filippenko \& Sargent 1993), so most of this gas would have to be
obscured at optical wavelengths in this interpretation.

\subsection{Synopsis of Absorption Results}

In summary, our data are consistent with the absorptions occurring in any of:
(i) a warm molecular region of density $\gtrsim 10^{9}$ cm$^{-3}$ and
ionization fraction several $\times$ 10$^{-5}$,
(ii) a hot atomic region of density $\sim 10^{5}$ cm$^{-3}$
and ionization fraction a few $\times$
10$^{-2}$ which co-exists with the 
masing molecular gas about 1 pc from the nuclear X-ray source, or
(iii) a combination of atomic and molecular regions, with the hot
atomic region dominating the free-free absorption and the molecular
region dominating the photoelectric absorption. 

\section{Conclusions}

We have found that the spectrum of the nuclear radio source in NGC 2639
turns over below $\sim$ 5 GHz and shown that this effect may result from
either synchrotron self-absorption or free-free absorption. If the latter
process is responsible, the implied value of $\int n_{e}^{2} dl$ may be
combined with the equivalent hydrogen column, $\int n_{H} dl$
(derived from the X-ray spectrum), to provide
the ionization fraction of the gas in terms of the electron
temperature and hydrogen density, assuming the same gas is responsible for
both kinds of absorption. The lower limit to the ionization fraction
obtained by assuming the absorptions occur in the masing 
molecular gas is comparable to the upper limit to the ionization fraction
derived theoretically by Neufeld, Maloney \& Conger (1994) for warm
molecular gas heated by a nuclear hard X-ray source, implying that
the molecular gas must be dense ($\gtrsim 10^{9}$ cm$^{-3}$) if it is to
provide the free-free absorption.
The
required ionization fraction is consistent with that expected in the atomic
phase at
5,000 -- 8,000 K which may co-exist with the molecular component.
If our line of sight passes through {\it both} phases, the hot atomic phase
would likely dominate the free-free absorption while molecular gas could
dominate the photoelectric absorption.
We rule out the hypothesis that both types of absorption occur in a fully
ionized gas at 10$^{4}$ K. 
Planned VLBA imaging of the H$_{2}$O megamaser will help distinguish between
these various possibilities.

We thank J. H. Krolik and D. A. Neufeld for comments on an early draft of the
manuscript
and P. R. Maloney for advice on the expected ionization fractions.
The
National Radio Astronomy Observatory is a facility
of the National Science Foundation,
operated under cooperative agreement by Associated
Universities, Inc.
This research was supported by NASA through grants NAG 53393 and NAG 81027,
by NSF through grant AST9527289 and by NATO through grant SA.5-2-05
(GRG. 960086) 318/96.

\vfil\eject

\begin{deluxetable}{lcccc}
\tablecolumns{4}
\tablewidth{0pc}
\tablecaption{Flux Densities, Beam Sizes (FWHM) and Position 
from VLBA Observations}
\tablehead{
& \multicolumn{3}{c}{Frequency (GHz)} \\
\cline{2-4} \\
& \colhead{1.7} & \colhead{5.0}&\colhead{15.3} \\ }
\startdata
Flux Density (mJy) & 6.9$\pm$0.6    & 47$\pm$4       & 40$\pm$3        \nl 
Beam Size (mas)    & 6.4$\times$4.6 & 2.0$\times$1.7 & 1.2$\times$0.75 \nl
                   &                &                &                 \nl
Flux Density (mJy) &                & 47$\pm$4       & 45$\pm$4        \nl
Beam Size (mas)    &                & 6.4$\times$4.6 & 6.4$\times$4.6  \nl
\enddata
\label{tab:comp-pos}
\tablenotetext{} {J2000 position of peak radio flux:
RA $=08^{h}$ $43^{m}$ $38^{s}.07788$,
Dec = $50^{\circ}$ $12'$ $20''.0044$ $\pm 0.9$
mas (absolute position
uncertainty).}
\end{deluxetable}

\clearpage

\begin{deluxetable}{llllll}
\tablecolumns{6}
\tablecaption{Models of the X-ray spectrum}
\tablehead{
Model$^{a}$ & $\Gamma^{b}$  &  N$_{H}$(soft)  & N$_{H}$(hard) &
$\chi^{2}$/d.o.f. & $\chi^{2}$/d.o.f.\\
      &               & \colhead(10$^{22}$ cm$^{-2}$) &
\colhead(10$^{22}$ cm$^{-2}$) & SB$^{c}$ & BB$^{c}$ \\}
\startdata
A       & 1.8 (1.4 - 2.4)$^{d}$ & 0 (0 - 0.19)$^{d}$ &
        & 56.5/45 & 55.3/45 \nl
B$^{e}$ & 2.4 (1.9 - 2.8)$^{d}$ &  0.03$^{f,g}$      & 42 (19 - 98)$^{d}$
			      & 44.2/44 & 45.6/44 \nl
C       & 1.9$^{g}$             &  0.03$^{f,g}$      & 53 (22 - 130)$^{d}$
					  & 47.7/45 & 48.5/45 \nl
D$^{h}$ & 1.9$^{g}$             &  0.03$^{f,g}$      &
		      & 46.7/45 & 42.0/45 \nl
\enddata
\tablenotetext{a} {Models are A: uniformly-absorbed power law;
B: power-law source partially covered
by N$_{H}$(hard) and fully covered by N$_{H}$(soft);
C: As model B, but with $\Gamma$ fixed at 1.9;
D: power law plus Gaussian (intended to represent Fe K$\alpha$ emission)
with Galactic absorption only.}
\tablenotetext{b} {Photon index.}
\tablenotetext{c} {SB and BB represent fits with the G3 background
taken from the source and blank-sky fields, respectively.}
\tablenotetext{d} {The values given represent the 90\% confidence ranges for
the two free parameters
$\Gamma$ and N$_{H}$ in models A and B, and the
90\% confidence range for the single free parameter
N$_{H}$ in model C. Errors
on $\Gamma$ and N$_{H}$ include the uncertainty in
the G3 background.}
\tablenotetext{e} {Cloud covering factor is 0.93 (90\% confidence range is
0.83 - 0.98).}
\tablenotetext{f} {Galactic hydrogen column density toward NGC 2639.}
\tablenotetext{g} {Fixed parameter.}
\tablenotetext{h} {The energy and equivalent width of the Gaussian are
5.5 (90\% confidence range is 5.0 - 5.7) keV and $\sim3$ keV, respectively.
The line width is fixed at $\sigma = 0.01$ keV.}
\tablenotetext{} {Note: the $0.6 - 10$ keV and
$2 - 10$ keV observed (absorbed) fluxes are 
$\simeq$ 6 $\times$ 10$^{-13}$ and $\simeq$
4 $\times$ 10$^{-13}$ erg cm$^{-2}$ s$^{-1}$, respectively. The
$0.6 - 10$ keV and
$2 - 10$ keV
intrinsic (unabsorbed) fluxes are
$\simeq$ 4 $\times$ 10$^{-12}$ and $\simeq$ 2 $\times$ 10$^{-12}$
erg cm$^{-2}$ s$^{-1}$, respectively.}
\end{deluxetable}

\clearpage

\figcaption[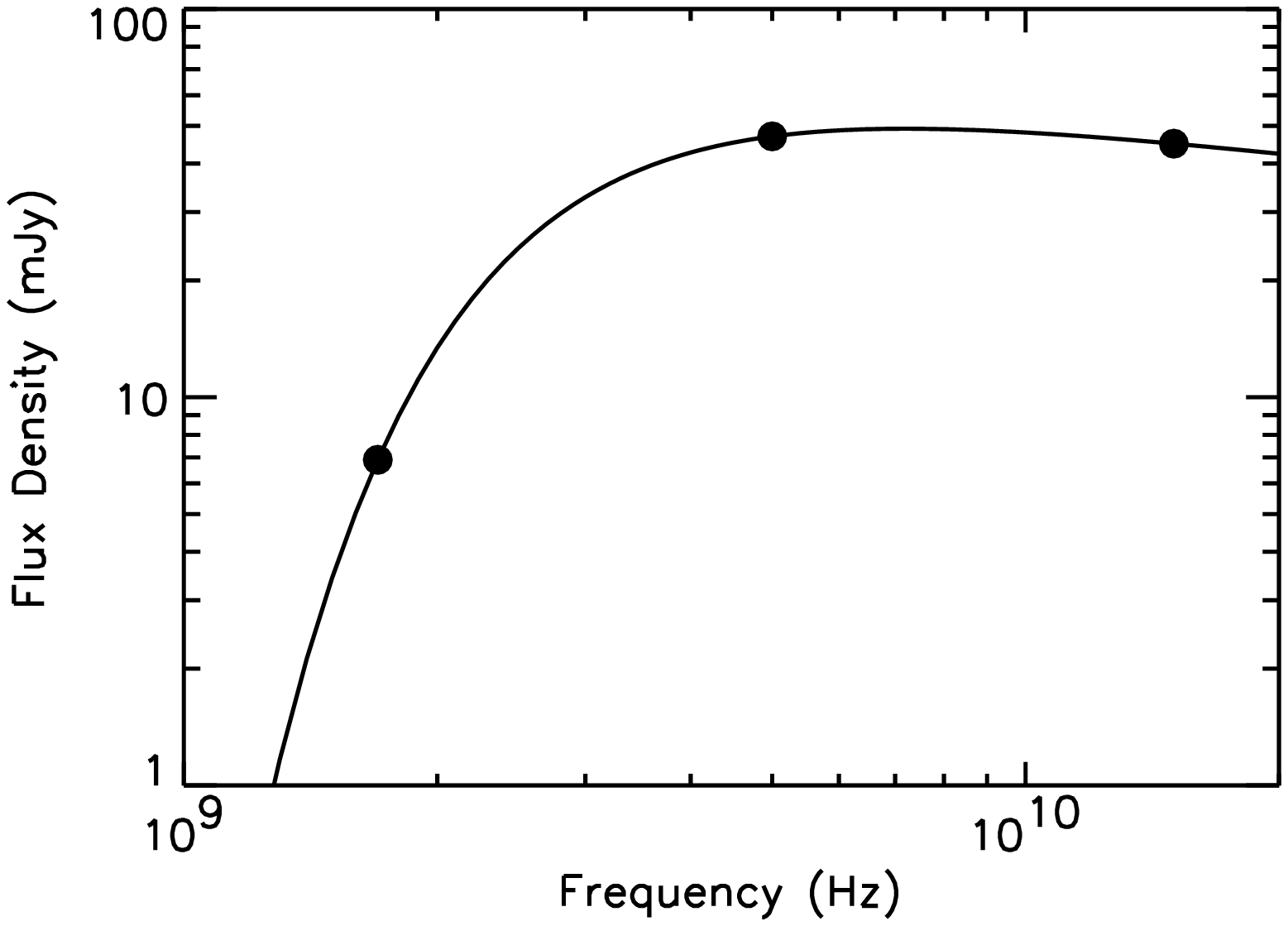]{The radio spectrum of the nuclear source in NGC 2639.
The points
are the VLBA
measurements, with the radii of the circles equal to the r.m.s.
uncertainties in flux density. The line represents a model comprising a source
with a
power-law spectrum
which is free-free absorbed by intervening ionized gas (see Section 3.1).
\label{Figure 1}}

\figcaption[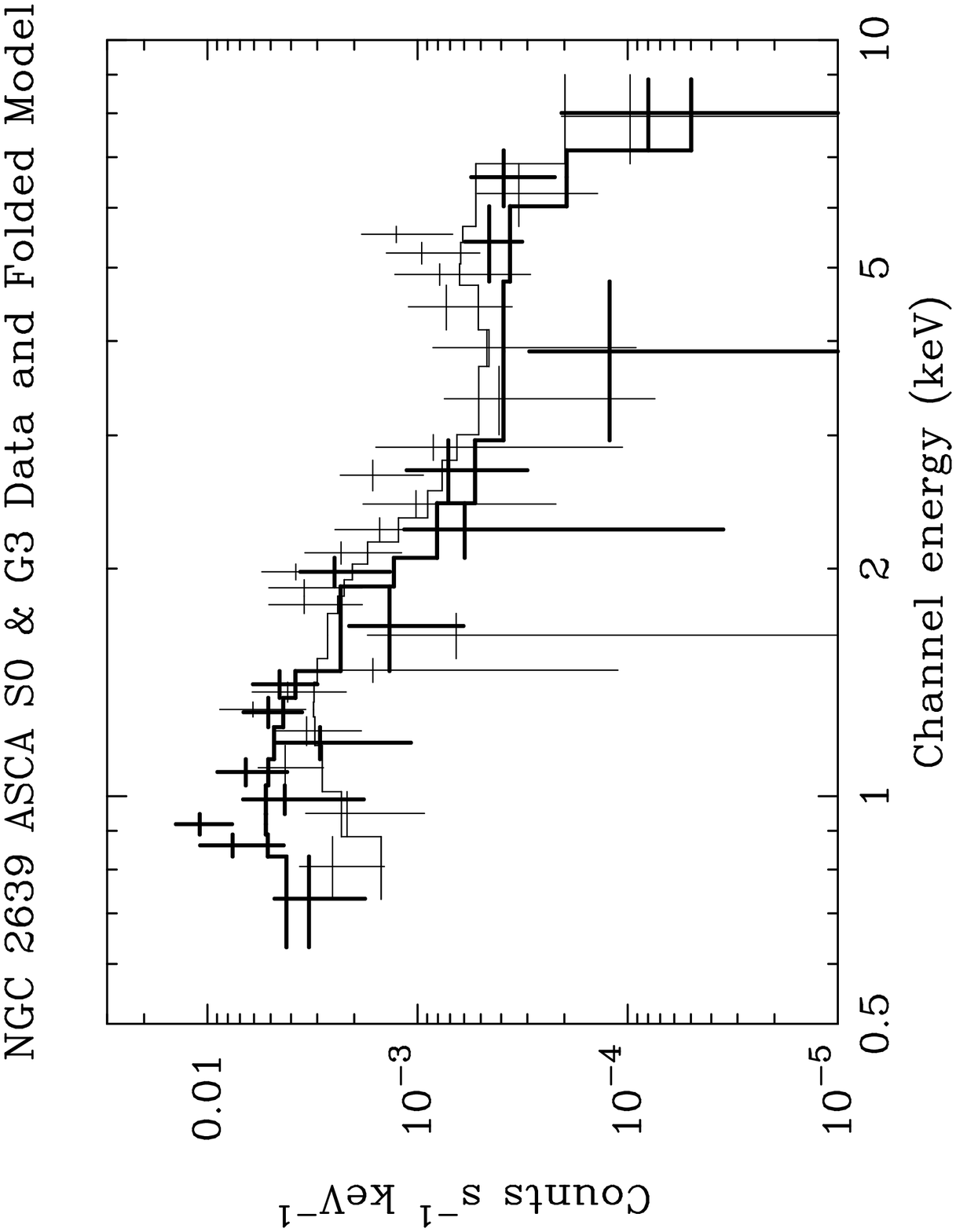]
{The observed ASCA spectra of NGC 2639
from detectors S0 and G3 along
with model B (see Table 2)
folded through the instrumental response.
Thick lines are S0 data and model; thin lines
are G3 data and model.
\label{Figure 2}}

\end{document}